\documentstyle[prl,aps]{revtex}

\begin{document}
\author{Jian-Qi Shen$^{1,2}$\footnote{E-mail address: jqshen@coer.zju.edu.cn}, Zhi-Chao Ruan$^1$,
and Sailing He$^{1,3}$\footnote{E-mail address: sailing@kth.se}}
\address{$^1$ Centre for Optical
and Electromagnetic Research, Joint Research Centre of Photonics
of the Royal Institute of Technology (Sweden) and Zhejiang
University, Zhejiang University, Hangzhou Yuquan 310027, P. R.
China\\
$^2$ Zhejiang Institute of Modern Physics and Department of
Physics, Zhejiang University, Hangzhou 310027, P. R. China\\
$^3$ Laboratory of Photonics and Microwave Engineering,
Department of Microelectronics and Information Technology,\\
Royal Institute of Technology, Electrum 229, SE-164 40 Kista,
Sweden}
\date{\today}
\title{Explicit formula for the external driving field used to eliminate
the decoherence\\ of a two-state system coupled to a noise field}

\maketitle

\begin{abstract}
The elimination of decoherence of a multiphoton two-state quantum
system by using an appropriate external driving field is
considered. The multiphoton process caused by the noise field has
a supersymmetric Lie algebraic structure. The time-evolution
equations for the off-diagonal elements of the density operator of
the two-state system are derived in the interaction picture. A
simple explicit formula is given for the time-dependent external
driving field used to eliminate the decoherence of the multiphoton
two-state system.
\\

{\it PACS:} 03.65.Yz, 42.50.Ct, 42.50.Gy, 32.80.Qk, 42.50.Hz

{\it Key words:} maintenance of coherence, decoherence, two-state
quantum system, multiphoton process, external driving field, noise
field \end{abstract}

\pacs{03.65.Yz, 42.50.Ct, 42.50.Gy}

\section{Introduction}
During the past decade, quantum computation, which utilizes the
superposition of many states of a quantum system, has attracted
extensive attention \cite{Shor,Cirac,Sackett,Barrett,Nielsen}.
Quantum computers can carry out computational tasks that are
intractable for conventional computers since a quantum computing
system composed of qubits (two-level quantum systems) has a
mysterious property of quantum coherence (including the
entanglement or quantum correlation), which has no counterpart in
the classical realm\cite{Nielsen}. However, because of the
existence of the {\it decoherence} it is not clear whether we can
finally build a truly practical and effective quantum
computer\cite{Shor}. Decoherence is essential to understanding how
classical physics emerges from quantum mechanics. It is such a
process by which a quantum-mechanical system that interacts with
its natural environment loses the characteristic properties that
distinguish quantum mechanics from classical physics. In other
words, decoherence arises from the unavoidable interaction between
the quantum systems and the noise fields, which are mostly
responsible for damping of the quantum coherence of systems such
as spin and two-state systems. Thus, if the decoherence occurs in
the computation, some qubits become entangled with the environment
and consequently the state of the quantum computer
collapses\cite{Shor}. Recently, much attention has been attracted
to the theoretical and experimental work for topological quantum
computation
\cite{Zanardi,Jones,Margolin,Zhu,Vedral,Wang2,Leibfried}. It is
well known that due to its topological and global
nature\cite{Berry}, the geometric phase of a spinning particle
will not be affected by the random fluctuation arising in the
evolution path\cite{Zanardi,Jones}. This means that the geometric
phase shift gate (should such exist) may be robust with respect to
certain types of operational errors. Since in a quantum computer
an error-tolerant quantum logic gate is particularly essential for
realizing a quantum information processor, the quantum computation
based on such a geometric phase shift gate will inevitably become
an ideal scenario to this purpose\cite{Wang2}. Indeed, in the
experiments performed at the end of last century, such conditional
geometric phase shift gates were realized through the nuclear
magnetic resonance (NMR) under an adiabatic
condition\cite{Cory,Gershenfeld,Jones2,Jones3}. The result of
quantum computation will be exact only when the adiabatic
condition can be fulfilled ({\it i.e.}, only when the
time-dependent parameters of the Hamiltonian of qubits vary
extremely slowly) \cite{Jones}. Once the adiabatic condition is
satisfied, we meet, however, the problem of quantum decoherence
caused by the inevitable interaction between the qubits (spin) and
the environments. The existence of such decoherence effects demand
that the quantum computation process (in which the maintenance of
coherence over a large number of states is important) must be
completed within the decoherence time. If not, the error caused by
the decoherence will inevitably increase. Thus, we need to change
the parameters of Hamiltonian rapidly. However, as discussed
before, such rapid changes may break down the adiabatic
requirement and leads to the nonadiabaticity error in the results
of quantum computation\cite{Jones}. Therefore, it seems that there
is inevitably a conflict between the adiabatic condition and the
requirement for removing the decoherence effects. In the
literature, there are several methods, such as error-avoiding
codes\cite{Zanardi2}, error-correcting codes\cite{Gottesman} and
decoherence-avoiding scheme \cite{Viola}, which can be used to
reduce the decoherence effects in qubits. In the
decoherence-avoiding scheme \cite{Viola}, the interaction between
the two-state quantum system and the environment (such as a noise
field, thermal reservoir and bath) is eliminated by using an
external controllable driving field. In the present paper, we
study the maintenance of coherence through the
decoherence-avoiding scheme in a two-state system. Note that in
order to consider the general interaction between the two-state
system and the noise field we treat the multiphoton process, {\it
i.e.}, the total number of photons created or annihilated in each
two-state transition process caused by a noise field is greater
than one.

\section{Interaction of a two-state system with the noise field and the driving field}

The interaction between the two-state system and the noise field
can be described by the supersymmetric multiphoton model
\cite{Sukumar,Kien,Lu2,Shen,Shen2}, and the Hamiltonian under the
rotating wave approximation can be written as \cite{Sukumar,Kien}

\begin{equation}
H=\frac{\omega _{0}}{2}\sigma _{z}+\omega
a^{\dagger}a+g(a^{\dagger })^{k}\sigma _{-}+g^{\ast }a^{k}\sigma
_{+}, \label{eq31}
\end{equation}
where the creation and annihilation operators $a^{\dagger }$ and
$a$ for the photons obey the commutation relation $\left[
a,a^{\dagger }\right] =1$, $\sigma _{\pm }$ and $\sigma _{z}$
denote the two-state system operators satisfying the commutation
relation $\left[ \sigma _{z},\sigma _{\pm }\right] =\pm 2\sigma
_{\pm }$, and $g$ {\it and} $ g^{\ast }$ are the multiphoton
coupling coefficients. The integer $k$ is the number of photons
created and annihilated in each two-state transition process, and
$\omega _{0}(t)$ {\it and} $\omega (t)$ the two-state transition
frequency and the mode frequency of photons, respectively.
Defining the following algebraic generators \cite{Lu2,Shen,Shen2}
\begin{eqnarray}
N &=&a^{\dagger }a+\frac{k-1}{2}\sigma _{z}+\frac{1}{2}=\left(
\begin{array}{cc}
a^{\dagger }a+\frac{k}{2} & 0 \\
0 & aa^{\dagger }-\frac{k}{2}
\end{array}
\right) ,\quad N^{^{\prime }}=\left(
\begin{array}{cc}
a^{k}(a^{\dagger })^{k} & 0 \\
0 & (a^{\dagger })^{k}a^{k}
\end{array}
\right) ,  \nonumber \\
Q &=&(a^{\dagger })^{k}\sigma _{-}=\left(
\begin{array}{cc}
0 & 0 \\
(a^{\dagger })^{k} & 0
\end{array}
\right) ,\quad Q^{\dagger }=a^{k}\sigma _{+}=\left(
\begin{array}{cc}
0 & a^{k} \\
0 & 0
\end{array}
\right),  \label{eq32}
\end{eqnarray}
one can easily show that $(N,N^{^{\prime }},Q,Q^{\dagger })$ form
a set of supersymmetric generators and possess the supersymmetric
Lie algebraic structure, {\it i.e.},

\begin{eqnarray}
Q^{2} &=&(Q^{\dagger })^{2}=0,\quad \left[ Q^{\dagger },Q\right]
=N^{^{\prime }}\sigma _{z},\quad \left[ N,N^{^{\prime }}\right] =0,\quad %
\left[ N,Q\right] =Q,  \nonumber \\
\left[ N,Q^{\dagger }\right] &=&-Q^{\dagger },\quad \left\{
Q^{\dagger },Q\right\} =N^{^{\prime }},\quad \left\{ Q,\sigma
_{z}\right\} =\left\{
Q^{\dagger },\sigma _{z}\right\} =0,  \nonumber \\
\left[ Q,\sigma _{z}\right] &=&2Q,\quad \left[ Q^{\dagger },\sigma _{z}%
\right] =-2Q^{\dagger },\quad \left( Q^{\dagger }-Q\right)
^{2}=-N^{^{\prime }},        \label{eq33}
\end{eqnarray}
where $\left\{ {}\right\} $ denotes the anticommuting bracket.

To eliminate the quantum decoherence, an external classical field
$E(t)$ is applied to the above system. The Hamiltonian for the
interaction between the two-state system and the driving field
$E(t)$ is given by
\begin{equation}
H_{\rm
E}=-\frac{1}{2}\left(dE\sigma_{+}+d^{\ast}E^{\ast}\sigma_{-}\right),
\end{equation}
where $d$ and $d^{\ast}$ denote the transition dipole moments of
this two-state quantum system. The total Hamiltonian is $H_{\rm
tot}=H+H_{\rm E}$. Note that here the environmental noise field
(weak) that is coupled to the two-state system is modelled by a
quantized electromagnetic field while the external driving field
(strong) is expressed by a set of classical quantities, $E$ and
$E^{\ast}$.

The Hamiltonian of the above multiphoton two-state system in the
interaction picture can be obtained by using the following unitary
transformation
\begin{equation}
V(t)=\exp\left[\frac{1}{i}\left(\frac{\omega _{0}}{2}\sigma
_{z}+\omega a^{\dagger }a\right)t\right].
\end{equation}
With the help of $H_{\rm I}(t)=V^{\dagger}(t)\left(H_{\rm
tot}-i\frac{\partial}{\partial t}\right)V(t)$, one can arrive at
\begin{equation}
H_{\rm I}(t)=g\exp(-i\delta t)Q+g^{\ast}\exp(i\delta
t)Q^{\dagger}-\frac{1}{2}\left(dEe^{i\omega_{0}t}\sigma_{+}+d^{\ast}E^{\ast}e^{-i\omega_{0}t}\sigma_{-}\right),
\end{equation}
which is the Hamiltonian in the interaction picture, where the
frequency detuning $\delta=k\omega-\omega _{0}$.

\section{Evolution of the off-diagonal density matrix elements}

In this section, we use the Markoff
approximation\cite{Louisell,Louisell2} to obtain the
time-evolution equation of the off-diagonal elements of the
density operator for this two-state quantum system driven by the
external classical field $E(t)$.

It is well known that the density operator of the two-state
system undergoing a multiphoton interaction satisfies the
following Liouville equation
\begin{equation}
i\frac{\partial \rho_{\rm I}(t)}{\partial t}=[H_{\rm I}(t),
\rho_{\rm I}(t)].   \label{Liouville}
\end{equation}
Let $\rho_{\rm s\rm I}(t)$ denote the reduced density operator of
the two-state system. It follows from Eq. (\ref{Liouville}) that
\begin{equation}
\dot{\rho}_{\rm s\rm I}(t)={\rm Tr}_{\rm r}\dot{\rho}_{\rm
I}(t)=-\int^{t}_{0}{\rm Tr}_{\rm r}[H_{\rm I}(t), [H_{\rm I}(t'),
\rho_{\rm I}(t')]]{\rm d}t'.                   \label{eq25}
\end{equation}
Here dot denotes the derivative with respect to time. If the
thermal reservoir is assumed to be quite large, it does not change
much during the time evolution process of the two-state system.
Then we have $\rho_{\rm I}(t)\simeq {\rho}_{\rm s\rm
I}(t){\rho}_{{\rm r}{\rm I}}(0)$, where the density operator of
the thermal reservoir is ${\rho}_{{\rm r}{\rm I}}(0)\simeq
\sum_{\bf k}\exp (-\beta H_{\bf k})/{\rm Tr}\exp(-\beta H_{\bf
k})$ with $\beta=1/(k_{\rm B}T)$, $H_{\bf k}=\omega_{\bf k}
a^{\dagger}_{\rm k}a_{\rm k}$. If we take $E(t)\simeq E(t')$,
$\rho_{\rm I}(t')\simeq \rho_{\rm I}(t)$ (Markoff
approximation)\cite{Louisell,Louisell2}, Eq. (\ref{eq25}) can be
rewritten as
\begin{equation}
\dot{\rho}_{\rm s\rm I}(t)=-\int^{t}_{0}{\rm Tr}_{\rm r}[H_{\rm
I}(t), [H_{\rm I}(t'), {\rho}_{\rm s\rm I}(t){\rho}_{{\rm r}{\rm
I}}(0)]]{\rm d}t',                              \label{eq26}
\end{equation}
where the integrand can be rewritten as
\begin{eqnarray}
{\rm Tr}_{\rm r}[H_{\rm I}(t), [H_{\rm I}(t'), {\rho}_{\rm s\rm
I}(t){\rho}_{{\rm r}{\rm I}}(0)]]&=&{\rm Tr}_{\rm r}[H_{\rm
I}(t)H_{\rm I}(t'){\rho}_{\rm s\rm I}(t){\rho}_{{\rm r}{\rm
I}}(0)-H_{\rm
I}(t){\rho}_{\rm s\rm I}(t){\rho}_{{\rm r}{\rm I}}(0)H_{\rm I}(t')                  \nonumber \\
& & -H_{\rm I}(t'){\rho}_{\rm s\rm I}(t){\rho}_{{\rm r}{\rm
I}}(0)H_{\rm I}(t)+{\rho}_{\rm s\rm I}(t){\rho}_{{\rm r}{\rm
I}}(0)H_{\rm I}(t')H_{\rm I}(t)].     \label{eq27}
\end{eqnarray}
After some tedious derivations, we obtain the explicit results for
the four terms on the right-hand side of Eq. (\ref{eq27}). The
first term can be rewritten as
\begin{eqnarray}
{\rm Tr}_{\rm r}[H_{\rm I}(t)H_{\rm I}(t'){\rho}_{\rm s\rm
I}(t){\rho}_{{\rm r}{\rm I}}(0)]&=&<H_{\rm I}(t)H_{\rm
I}(t'){\rho}_{\rm s\rm I}(t)>_{\rm R}        \nonumber \\
&=&g^{\ast}g\left\{\exp[i\delta(t'-t)]<QQ^{\dagger}\rho_{\rm s\rm
I}(t)>_{\rm R} +\exp[-i\delta(t'-t)]<Q^{\dagger}Q\rho_{\rm s\rm
I}(t)>_{\rm R}\right\}            \nonumber \\
& &
+\frac{1}{4}d^{\ast}dE^{\ast}(t)E(t)\exp[i\omega_{0}(t-t')]{\rho}_{\rm
s\rm I}(t), \label{eq271}
\end{eqnarray}
where the relation $Q^{2}=(Q^{\dagger})^{2}=0$ has been applied to
the calculation. Here the subscript ${\rm R}$ means the average
over the states of the thermal reservoir (noise field). The rest
of the terms on the right-hand side of Eq. (\ref{eq27}) are given
by
\begin{eqnarray}
& & -{\rm Tr}_{\rm r}[H_{\rm I}(t){\rho}_{\rm s\rm
I}(t){\rho}_{{\rm r}{\rm I}}(0)H_{\rm I}(t')]=-<H_{\rm
I}(t){\rho}_{\rm s\rm I}(t)H_{\rm
I}(t')>_{\rm R}                                                      \nonumber \\
&=& -g^{2}\exp[-i\delta (t+t')]<Q{\rho}_{\rm s\rm I}(t)Q>_{\rm R}
-(g^{\ast})^{2}\exp[i\delta (t+t')]<Q^{\dagger}{\rho}_{\rm s\rm I}(t)Q^{\dagger}>_{\rm R}      \nonumber \\
& & -g^{\ast}g\left\{\exp[i\delta(t'-t)]<Q{\rho}_{\rm s\rm
I}(t)Q^{\dagger}>_{\rm
R}+\exp[-i\delta(t'-t)]<Q^{\dagger}{\rho}_{\rm s\rm I}(t)Q>_{\rm
R}\right\}     \nonumber \\
& &
-\frac{d^{2}}{4}E^{2}(t)\exp[i\omega_{0}(t+t')]\sigma_{+}{\rho}_{\rm
s\rm I}(t)\sigma_{+}-\frac{\left(d^{\ast}\right)
^{2}}{4}\left(E^{\ast}\right)^{2}(t)\exp[-i\omega_{0}(t+t')]\sigma_{-}{\rho}_{\rm
s\rm I}(t)\sigma_{-}
  \nonumber \\
& &
-\frac{d^{\ast}d}{4}\left\{E^{\ast}(t)E(t)\exp[i\omega_{0}(t-t')]\sigma_{+}{\rho}_{\rm
s\rm
I}(t)\sigma_{-}+E^{\ast}(t)E(t)\exp[-i\omega_{0}(t-t')]\sigma_{-}{\rho}_{\rm
s\rm I}(t)\sigma_{+}\right\},
\end{eqnarray}

\begin{eqnarray}
&-&{\rm Tr}_{\rm r}[H_{\rm I}(t'){\rho}_{\rm s\rm
I}(t){\rho}_{{\rm r}{\rm I}}(0)H_{\rm I}(t)]=-g^{2}\exp[-i\delta
(t+t')]<Q{\rho}_{\rm s\rm I}(t)Q>_{\rm R}
-(g^{\ast})^{2}\exp[i\delta (t+t')]<Q^{\dagger}{\rho}_{\rm s\rm I}(t)Q^{\dagger}>_{\rm R}      \nonumber \\
& & -g^{\ast}g\left\{\exp[-i\delta(t'-t)]<Q{\rho}_{\rm s\rm
I}(t)Q^{\dagger}>_{\rm
R}+\exp[i\delta(t'-t)]<Q^{\dagger}{\rho}_{\rm s\rm I}(t)Q>_{\rm
R}\right\}    \nonumber \\
& &
-\frac{d^{2}}{4}E^{2}(t)\exp[i\omega_{0}(t+t')]\sigma_{+}{\rho}_{\rm
s\rm I}(t)\sigma_{+}-\frac{\left(d^{\ast}\right)
^{2}}{4}\left(E^{\ast}\right)^{2}(t)\exp[-i\omega_{0}(t+t')]\sigma_{-}{\rho}_{\rm
s\rm I}(t)\sigma_{-}
  \nonumber \\
& &
-\frac{d^{\ast}d}{4}\left\{E^{\ast}(t)E(t)\exp[i\omega_{0}(t'-t)]\sigma_{+}{\rho}_{\rm
s\rm
I}(t)\sigma_{-}+E^{\ast}(t)E(t)\exp[-i\omega_{0}(t'-t)]\sigma_{-}{\rho}_{\rm
s\rm I}(t)\sigma_{+}\right\}
\end{eqnarray}
and
\begin{eqnarray}
{\rm Tr}_{\rm r}[{\rho}_{\rm s\rm I}(t){\rho}_{{\rm r}{\rm
I}}(0)H_{\rm I}(t')H_{\rm
I}(t)]&=&g^{\ast}g\left\{\exp[-i\delta(t'-t)]<\rho_{\rm s\rm
I}(t)QQ^{\dagger}>_{\rm R} +\exp[i\delta(t'-t)]<\rho_{\rm s\rm
I}(t)Q^{\dagger}Q>_{\rm R}\right\}          \nonumber  \\
& &
+\frac{d^{\ast}d}{4}E^{\ast}(t)E(t)\exp[-i\omega_{0}(t-t')]{\rho}_{\rm
s\rm I}(t).              \label{eq274}
\end{eqnarray}
Thus, the integrand in Eq. (\ref{eq26}) can be rewritten in the
following form
\begin{equation}
{\rm Tr}_{\rm r}[H_{\rm I}(t), [H_{\rm I}(t'), {\rho}_{\rm s\rm
I}(t){\rho}_{{\rm r}{\rm I}}(0)]]={\mathcal T}_{1}+{\mathcal
T}_{2}+{\mathcal T}_{3}+{\mathcal T}_{4},   \label{4tau}
\end{equation}
where
\begin{eqnarray}
{\mathcal
T}_{1}&=&g^{\ast}g\left\{\exp[i\delta(t'-t)]<(a^{\dagger})^{k}a^{k}>_{\rm
R}\sigma_{-}\sigma_{+}+\exp[-i\delta(t'-t)]<a^{k}(a^{\dagger})^{k}>_{\rm
R}\sigma_{+}\sigma_{-}\right\}{\rho}_{\rm s\rm
I}(t)                           \nonumber\\
& & +g^{\ast}g{\rho}_{\rm s\rm
I}(t)\left\{\exp[-i\delta(t'-t)]<(a^{\dagger})^{k}a^{k}>_{\rm
R}\sigma_{-}\sigma_{+}+\exp[i\delta(t'-t)]<a^{k}(a^{\dagger})^{k}>_{\rm
R}\sigma_{+}\sigma_{-}\right\},
\nonumber \\
{\mathcal T}_{2}&=& -2\left\{g^{2}\exp[-i\delta
(t+t')]<Q{\rho}_{\rm s\rm I}(t)Q>_{\rm
R}+(g^{\ast})^{2}\exp[i\delta (t+t')]<Q^{\dagger}{\rho}_{\rm s\rm
I}(t)Q^{\dagger}>_{\rm R}\right\},
\nonumber \\
{\mathcal T}_{3}&=&-g^{\ast}g\left\{\exp[i\delta
(t'-t)]+\exp[-i\delta (t'-t)]\right\}\left(<Q{\rho}_{\rm s\rm
I}(t)Q^{\dagger}>_{\rm R}+<Q^{\dagger}{\rho}_{\rm s\rm
I}(t)Q>_{\rm R}\right),
\nonumber \\
{\mathcal
T}_{4}&=&\frac{d^{\ast}d}{4}E^{\ast}(t)E(t)\left\{\exp[i\omega_{0}(t-t')]+\exp[-i\omega_{0}(t-t')]\right\}{\rho}_{\rm
s\rm
I}(t)               \nonumber \\
& &
-\frac{d^{2}}{2}E^{2}(t)\exp[i\omega_{0}(t+t')]\sigma_{+}{\rho}_{\rm
s\rm I}(t)\sigma_{+}-\frac{\left(d^{\ast}\right)
^{2}}{2}\left(E^{\ast}\right)^{2}(t)\exp[-i\omega_{0}(t+t')]\sigma_{-}{\rho}_{\rm
s\rm I}(t)\sigma_{-}
 \nonumber \\
& &
-\frac{d^{\ast}d}{4}E^{\ast}(t)E(t)\left\{\exp[i\omega_{0}(t-t')]+\exp[-i\omega_{0}(t-t')]\right\}\left[\sigma_{+}{\rho}_{\rm
s\rm I}(t)\sigma_{-}+\sigma_{-}{\rho}_{\rm s\rm
I}(t)\sigma_{+}\right].
\end{eqnarray}

Assume that the reservoir is in a multiphoton state $\left|
m\right\rangle$, where the occupation number $m \ge k$.  Using the
relations $a^{k}(a^{\dagger })^{k}\left| m\right\rangle
=\frac{(m+k)!}{ m!}\left| m\right\rangle $,
$<a^{k}(a^{\dagger})^{k}>_{\rm R}=\frac{(m+k)!}{ m!}$,
$<(a^{\dagger})^{k}a^{k}>_{\rm R}=\frac{m!}{ (m-k)!}$,
$<(a^{\dagger})^{k}(a^{\dagger})^{k}>_{\rm R}=0$,
$\sigma_{+}|+\rangle=\sigma_{-}|-\rangle=0$ and
 $\langle+|\sigma_{-}=\langle-|\sigma_{+}=0 $, one can obtain
\begin{eqnarray}
\langle -|{\mathcal T}_{1}|+\rangle &=&
g^{\ast}g\exp[i\delta(t'-t)]\left[\frac{(m+k)!}{
m!}+\frac{m!}{(m-k)!}\right]\langle -|{\rho}_{\rm s\rm
I}(t)|+\rangle,
 \nonumber   \\
\langle -|{\mathcal T}_{2}|+\rangle&=& 0,
\nonumber   \\
 \langle -|{\mathcal
T}_{3}|+\rangle&=&-g^{\ast}g\left\{\exp[i\delta
(t'-t)]+\exp[-i\delta (t'-t)]\right\}\frac{(m+k)!}{m!}
\langle+|{\rho}_{\rm s\rm I}(t)|-\rangle,                      \nonumber   \\
 \langle -|{\mathcal
T}_{4}|+\rangle &=&
\frac{d^{\ast}d}{4}E^{\ast}(t)E(t)\left\{\exp[i\omega_{0}(t-t')]+\exp[-i\omega_{0}(t-t')]\right\}\langle -|{\rho}_{\rm s\rm I}(t)|+\rangle               \nonumber \\
& & -\frac{\left(d^{\ast}\right)
^{2}}{2}\left[E^{\ast}(t)\right]^{2}\exp[-i\omega_{0}(t+t')]\langle
+|{\rho}_{\rm s\rm I}(t)|-\rangle,   \label{tau4}
\end{eqnarray}
where the relations $\langle -|\sigma_{+}{\rho}_{\rm s\rm
I}(t)\sigma_{-}|+\rangle=\langle -|\sigma_{-}{\rho}_{\rm s\rm
I}(t)\sigma_{+}|+\rangle=0$ has been inserted.

Define ${\rho}_{-+}(t)=\langle -|{\rho}_{\rm s\rm I}(t)|+\rangle$,
${\rho}_{+-}(t)=\langle +|{\rho}_{\rm s\rm I}(t)|-\rangle$ and
$\dot{{\rho}}_{-+}(t)=\langle -|\dot{{\rho}}_{\rm s\rm
I}(t)|+\rangle$.  It follows from Eq. (\ref{eq26}) that
\begin{eqnarray}
\dot{{\rho}}_{-+}(t)=-\int^{t}_{0}{\rm Tr}_{\rm r}\langle
-|[H_{\rm I}(t), [H_{\rm I}(t'), {\rho}_{\rm s\rm
I}(t){\rho}_{{\rm r}{\rm I}}(0)]]|+\rangle {\rm d}t'.
\label{eq3160}
\end{eqnarray}
Substituting Eqs. (\ref{4tau}) and (\ref{tau4}) into the above
equation and neglecting the rapidly oscillating term ({\it i.e.},
the last term) in $\langle -|{\mathcal T}_{4}|+\rangle$, we obtain
\begin{eqnarray}
\dot{{\rho}}_{-+}(t)=c_{1}(t){\rho}_{-+}(t)-c_{2}(t){\rho}_{+-}(t),
 \label{eq316}
\end{eqnarray}
where the time-dependent coefficients $c_{1}(t)$ and $c_{2}(t)$
are given by
\begin{equation}
\left\{
\begin{array}{ll}
&   c_{1}(t)=-g^{\ast}g\left[\frac{(m+k)!}{
m!}+\frac{m!}{(m-k)!}\right]\frac{1-\exp(-i\delta t
)}{i\delta}+\frac{d^{\ast}d}{2\omega_{0}}E^{\ast}(t)E(t)\sin
\omega_{0}t,
  \\
&
   c_{2}(t)=-g^{\ast}g\frac{(m+k)!}{m!}\left[\frac{1-\exp(-i\delta t
)}{i\delta}+\frac{1-\exp(i\delta t )}{-i\delta}\right].
\end{array}
\right.      \label{eqeq}
\end{equation}
Note that the coefficient $c_{1}(t)$ is complex and $c_{2}(t)$ has
a real value.

The complex conjugation of Eq. (\ref{eq316}) gives
\begin{equation}
\dot{{\rho}}_{+-}(t)=c^{\ast}_{1}(t){\rho}_{+-}(t)-c_{2}(t){\rho}_{-+}(t).
\label{eq317}
\end{equation}
Eqs. (\ref{eq316}) and (\ref{eq317}) are the time-dependent
equations of the off-diagonal elements of the density operator for
this multiphoton two-state quantum system.

\section{Elimination of decoherence in the multiphoton
 two-state system}

In order to maintain the coherence of the two-state system
interacting with the noise field, we should analyze Eqs.
(\ref{eq316}) and (\ref{eq317}), which govern the time evolution
of the off-diagonal matrix elements of the density operator. It is
obvious that in the presence of noise field the decoherence can be
eliminated if $\dot{{\rho}}_{+-}(t) = \dot{{\rho}}_{-+}(t)
\rightarrow 0$ is satisfied. This leads to the requirement
\begin{equation}
c_{1}^{\ast}c_{1}-c_{2}^{2}=0.   \label{significance}
\end{equation}

In what follows, we will solve $E(t)$. For convenience, we first
define
\begin{equation}
 \alpha(t)=-g^{\ast}g\left[\frac{(m+k)!}{
m!}+\frac{m!}{(m-k)!}\right]\frac{1-\exp(-i\delta t )}{i\delta},
\quad  \beta(t)=\frac{d^{\ast}d}{2\omega_{0}}\sin \omega_{0}t,
\quad x(t)=E^{\ast}(t)E(t),
\end{equation}
and then rewrite the coefficient $c_{1}(t)$ as follows
\begin{equation}
c_{1}(t)=\alpha(t)+ \beta(t)x(t).
\end{equation}
According to Eq. (\ref{significance}), one obtains
$(\alpha^{\ast}+ \beta x)(\alpha+ \beta x)-c_{2}^{2}=0$, which can
be rewritten as
\begin{equation}
x^{2}+\frac{\alpha^{\ast}+\alpha}{\beta}x+\frac{\alpha^{\ast}\alpha-c_{2}^{2}}{\beta^{2}}=0.
\label{signif}
\end{equation}
The two roots of the above quadratic equation are of the form
\begin{equation}
x_{\pm}=\frac{-\frac{\alpha+\alpha^{\ast}}{\beta}\pm
\frac{1}{|\beta|}\sqrt{\left(\alpha-\alpha^{\ast}\right)^{2}+4c^{2}_{2}}}{2},
\label{eqEE}
\end{equation}
where
\begin{equation}
\alpha+\alpha^{\ast}=-g^{\ast}g\left[\frac{(m+k)!}{
m!}+\frac{m!}{(m-k)!}\right]\frac{2\sin \delta t}{\delta}, \quad
\alpha-\alpha^{\ast}=-g^{\ast}g\left[\frac{(m+k)!}{
m!}+\frac{m!}{(m-k)!}\right]\frac{2(1-\cos \delta t)}{i\delta}.
\label{alpha}
\end{equation}
If we apply an external time-dependent driving field with
intensity given by Eq. (\ref{eqEE}), in principle the decoherence
of the two-state quantum system interacting with the environment
(noise field) can be eliminated. Fig. 1 shows the external
time-dependent driving field required for the elimination of the
decoherence of the multiphoton two-state system with various $k$
(the number of photons in each transition process) and $\delta=0$
(without detuning). It follows from expressions (\ref{eqeq}) and
(\ref{alpha}) that all the parameters $c_{2}$,
$\alpha+\alpha^{\ast}$ and $\alpha-\alpha^{\ast}$ are inversely
proportional to the frequency detuning $\delta$. It is thus shown
that according to Eqs. (\ref{eq316})-(\ref{eq317}), the two-state
system may not be subject to the quantum decoherence significantly
in the case of large detuning. In what follows, as an illustrative
example, we will consider only the maintenance of the coherence
for the case of small detuning. Fig. 2 demonstrates the required
time-dependent expression for the external driving field when the
frequency detuning is $\delta =0.01 \omega _0$. It should be noted
that in some time ranges ({\it e.g.}, $2.2\times
10^{-10}<t<2.5\times 10^{-10}$ s for the case of $k=1$) in Fig. 2,
both the field intensities $x_{+}(t)$ and $x_{-}(t)$ obtained in
(\ref{signif}) are negative, {\it i.e.}, $E^{\ast}(t)E(t)<0$,
which has no physical meanings. So, in the time ranges where
$E^{\ast}(t)E(t)<0$, what one should do is to set $E(t)=0$
experimentally. This, however, means that if we utilize the
external driving field illustrated in Fig. 2, the coherence cannot
be maintained completely after $t=2.2\times 10^{-10}$ s (for the
case of $k=1$). To overcome such a difficulty, here we will
suggest a potential scheme of cyclically driving field:
specifically, if we choose the evolutional behavior of the
external driving field of the time range $[0,T]$ ($T=2.2\times
10^{-10}$ s) and apply it to any ranges $[nT, (n+1)T]$, namely,
$E(nT+t)=E(t)$, where $t\in [0,T]$, the quantum decoherence of the
above two-state system undergoing a multiphoton interaction caused
by the environmental noise field will be greatly reduced.

To summarize, the multiphoton process, which possesses a
supersymmetric Lie algebraic structure, is caused by the
interaction between the two-state system and the noise field of
the environment. This will lead to the decoherence of the
two-state quantum system. In the present paper, we have considered
the reduction of decoherence by using an appropriate external
time-dependent driving field. We have used the Markoff
approximation to derive the equations governing the time-evolution
behavior of the off-diagonal elements of the two-state density
operator in the interaction picture. A simple explicit formula for
the intensity of the external driving field has been given for the
elimination of the decoherence of the multiphoton two-state
system.
\\ \\
\textbf{Acknowledgements} This work is supported by the National
Natural Science Foundation of China under Project Nos. $90101024$
and $60378037$.

\newpage

\section*{Figure captions}

Fig. 1. The external time-dependent driving field required for the
elimination of the decoherence of the two-state system coupled to
a noise field. Here $m=100$, $\omega_{0}=1.0\times 10^{11}$
s$^{-1}$ and the frequency detuning $\delta$ vanishes.
\\ \\

Fig. 2. The external time-dependent driving field required for the
elimination of the decoherence of the two-state system coupled to
a noise field. Here $m=100$, $\omega_{0}=1.0\times 10^{11}$
s$^{-1}$ and the frequency detuning $\delta=0.01\omega_{0}$.

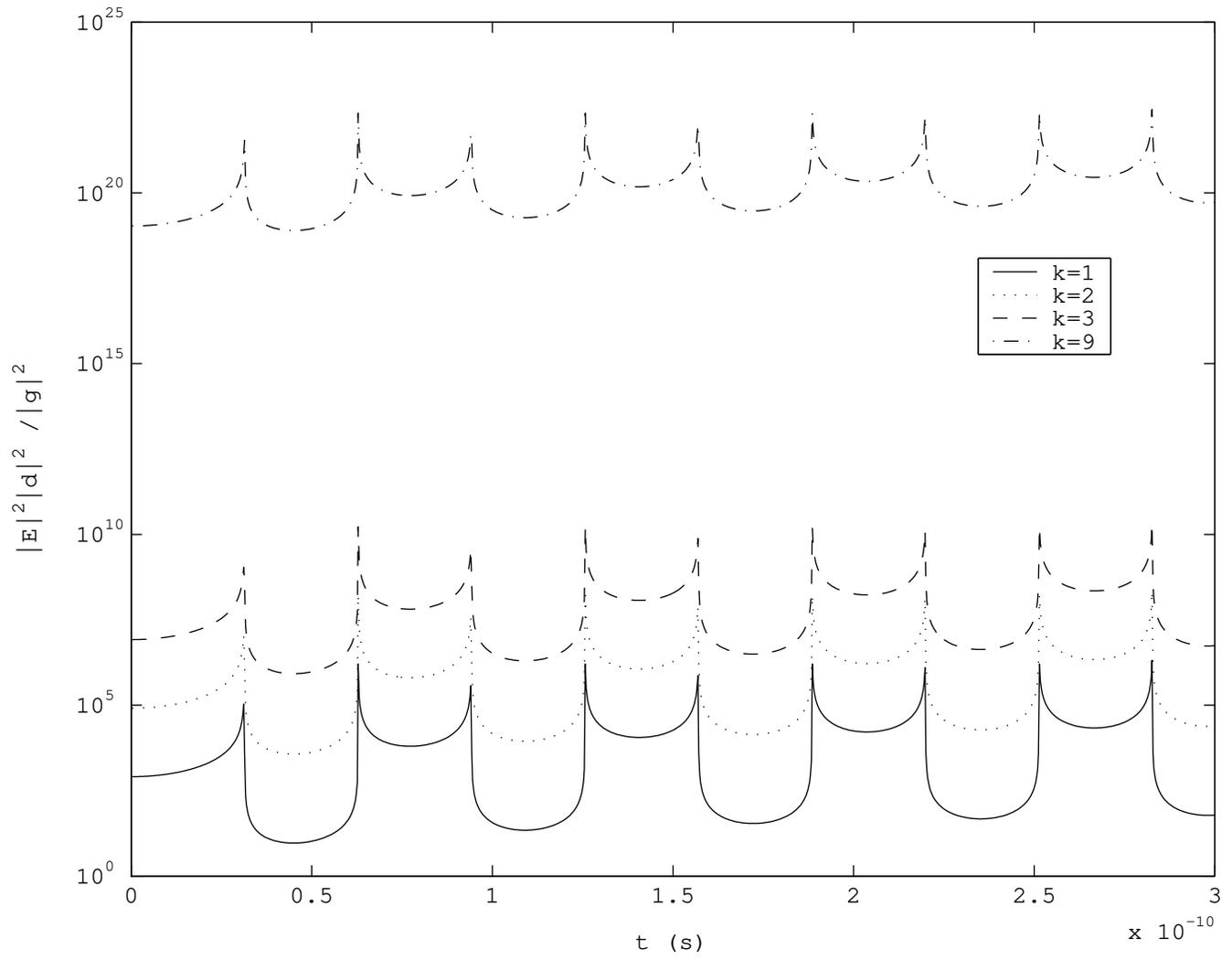
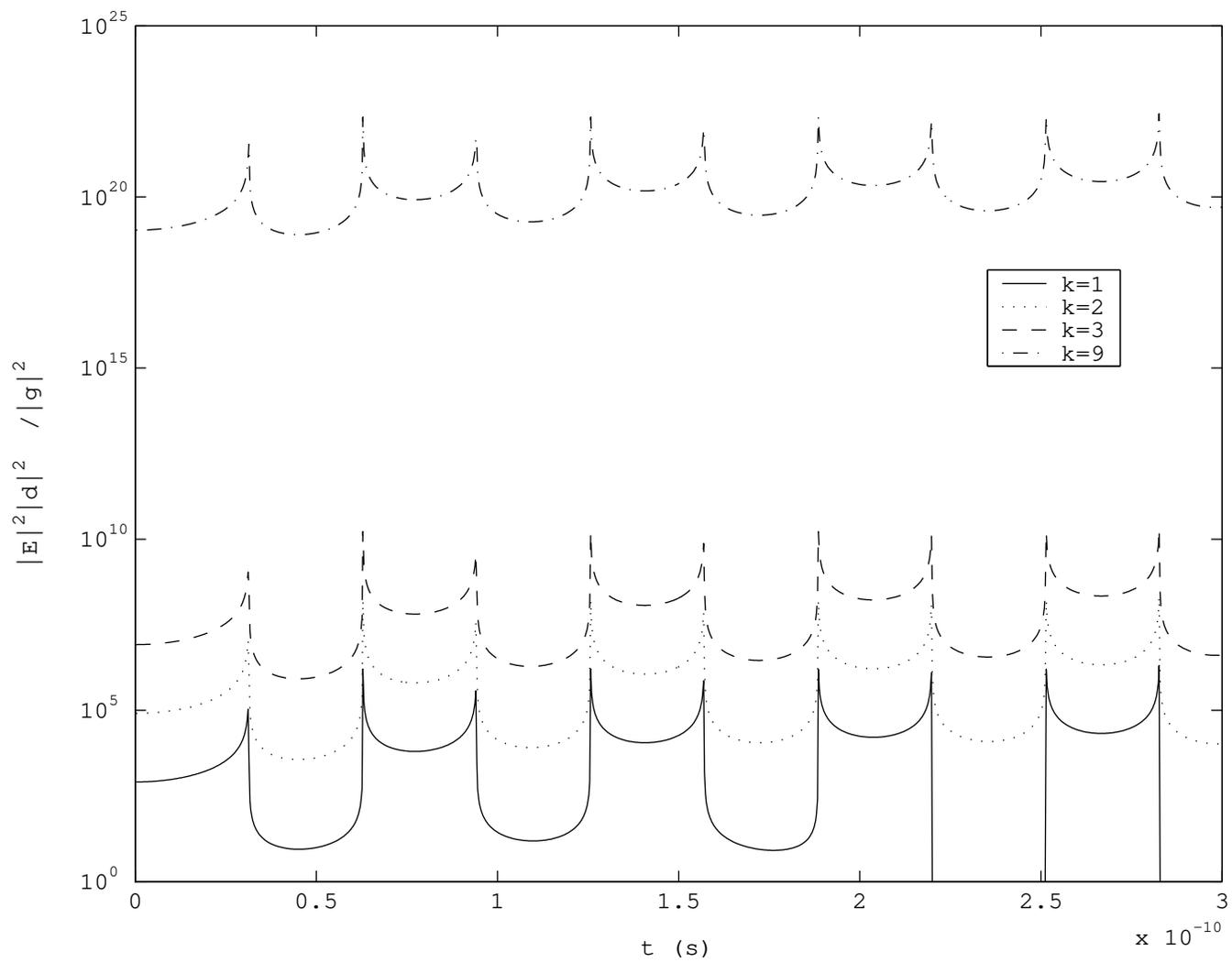
\end{document}